\documentclass[preprint,aps,prd,floatfix,nofootinbib,10pt]{revtex4-2}
\pdfoutput=1
\usepackage{graphicx}
\usepackage{amsfonts}
\usepackage{nicefrac}
\usepackage{xcolor}
\usepackage{natbib}
\usepackage{braket}
\usepackage{hyperref}

\usepackage{amsmath}
\usepackage{rotating}
\usepackage{amssymb,amsthm}
\usepackage{soul}

\usepackage{xcolor}
\usepackage[normalem]{ulem}

\usepackage{latexsym}
\usepackage{graphics}

\usepackage{amsfonts}
\usepackage{amsmath}
\usepackage{rotating}
\usepackage{amssymb}

\usepackage{amsmath}
\usepackage{rotating}
\usepackage{amssymb}
\usepackage{soul}

\usepackage{xcolor}

\usepackage{xcolor}
\def\beq{\begin{eqnarray}}  
	\def\eeq{\end{eqnarray}}

\def\beq{\begin{eqnarray}}  
	\def\eeq{\end{eqnarray}}

\begin{document}

\title{Electromagnetic Sources Teleparallel Robertson--Walker $F(T)$-Gravity Solutions}
\author{A. Landry}
\email{a.landry@dal.ca}
\affiliation{Department of Mathematics and Statistics, Dalhousie University, Halifax, Nova Scotia, Canada, B3H 3J5}

\begin{abstract}

{We investigate the teleparallel Robertson--Walker (TRW) $F(T)$-gravity solutions for a cosmological electromagnetic source in the current paper. We use and solve the TRW $F(T)$-gravity field equations (FEs) for each value of the $k$-parameter $(-1,\,0,\,+1)$ and the electromagnetic equivalent of the equation of state (EoS), leading to new teleparallel $F(T)$ solutions. For the $k=0$ cosmological case, we find new teleparallel $F(T)$ solutions for any scale factor $n$. For $k=\pm 1$ cosmological cases, we find exact and far-future approximated new teleparallel $F(T)$ solutions for slow, linear, fast and infinitely fast universe expansion summarized by analytical functions. All the new solutions are relevant for future cosmological applications, implying any electromagnetic source processes, such as the cosmological plasma models.}


\end{abstract}

	\maketitle
\tableofcontents

\newpage


\section{Introduction}\label{sec1}

The teleparallel $F(T)$-gravity is a frame-based alternative theory to general relativity (GR) defined in terms of a coframe/spin-connection pair ({${\bf h}^a$}--$\omega^a_{~bc}$ 
 pair) \cite{Lucas_Obukhov_Pereira2009,Aldrovandi_Pereira2013,Bahamonde:2021gfp,Krssak:2018ywd,MCH,Coley:2019zld,Krssak_Pereira2015}. The~two last quantities define the torsion tensor $T^a_{~bc}$ and torsion scalar $T$. We remind you that GR is defined by the metric $g_{\mu\nu}$ and the spacetime curvatures $R^a_{~b\mu\nu}$, $R_{\mu\nu}$, and $R$. We can determine the symmetries for any independent {${\bf h}^a$--$\omega^a_{~bc}$ pairs,} and then spacetime curvature and torsion are defined as geometric objects~\cite{MCH,Coley:2019zld,Krssak_Saridakis2015,Krssak_Pereira2015,Aldrovandi_Pereira2013,olver1995equivalence}. Any geometry described by such a pair {is a null curvature and non-metricity ($R^a_{~b\mu\nu}=0$ and $Q_{a\mu\nu}=0$ conditions), and~is considered as a teleparallel gauge-invariant geometry (for any gauge metric $g_{ab}$). The~fundamental pairs will satisfy two Lie derivative-based relations, and by using the Cartan--Karlhede algorithm, we can solve the} two fundamental equations for any teleparallel geometry. For~a pure teleparallel $F(T)$-gravity {solution, we solve the $R^a_{~b\mu\nu}=0$ condition and find a Lorentz transformation-based definition of $\omega^a_{~b\mu}$.} There is a direct equivalent to GR in teleparallel gravity: the teleparallel equivalent to GR (TEGR) generalizing to the teleparallel $F(T)$-gravity~\cite{Aldrovandi_Pereira2013,Krssak_Saridakis2015,Ferraro:2006jd,Ferraro:2008ey,Linder:2010py}. All the {previous considerations are also valid for the new general relativity (NGR) (refs.~\cite{kayashi,beltranngr,bahamondengr} and refs. therein), and the~symmetric teleparallel $F(Q)$ gravity (refs.~\cite{heisenberg1,heisenberg2,faithman1,hohmannfq} and refs. therein) and other theories like $F(T,Q)$-type, $F(R,Q)$-type, $F(R,T)$-type, and~other ones} (refs.~\cite{jimeneztrinity,nakayama,ftqgravity,frtspecial,frttheory,myrzakulov1,myrzakulov2,myrzakulov3,myrzakulov4,myrzakulov5} and refs. therein). {Therefore, the~current study will be restricted to the pure} teleparallel $F(T)$-gravity~framework.

There are a {huge amount of papers on spherically symmetric teleparallel $F(T)$-gravity solutions using a number of approaches, energy--momentum sources, and proceeding by} various purposes~\cite{golov1,golov2,golov3,debenedictis,SSpaper,TdSpaper,nonvacSSpaper,nonvacKSpaper,staticscalarfieldSS,scalarfieldKS,roberthudsonSSpaper,coleylandrygholami,scalarfieldTRW,baha1,bahagolov1,awad1,baha6,nashed5,pfeifer2,elhanafy1,benedictis3,baha10,baha4,ruggiero2,sahoo1,sahoo2,calza}. There is a special class of teleparallel spacetime in which the field equations (FEs) are purely symmetric: the teleparallel Robertson--Walker (TRW) spacetime~\cite{preprint,coleylandrygholami,scalarfieldTRW,ChaplyginpolyTRW}. The~TRW spacetime is defined in terms of the {cosmological $k$-parameter, where $k=0$ is a flat spacetime and $k=\pm 1$ are, respectively, positive and negative space curvature~\cite{aldrovandi2003,bounce,Capozz,inflat}.} A TRW geometry is described by a $G_6$ Lie algebra group, where the $4$th to $6$th Killing Vectors (KVs) are {specific} to this spacetime~\cite{preprint,coleylandrygholami}. The~main consequence of additional KVs is the trivial antisymmetric parts of FEs. The~{TRW spacetime structure is also defined} for a teleparallel $F(T,B)$ gravity extension~\cite{FTBcosmogholamilandry,HJKP2018,Cai_2015,dixit,ftbcosmo3}. We found teleparallel $F(T)$ and $F(T,B)$ solutions for perfect fluid (PF) and scalar field (SF) sources. The~SF-based teleparallel $F(T)$ and $F(T,B)$ solutions are scalar potential independent and only SF dependent~\cite{scalarfieldTRW,FTBcosmogholamilandry}. There are non-linear fluid teleparallel $F(T)$ solutions for polytropic and Chaplygin fluids leading to similar results~\cite{ChaplyginpolyTRW}. But~there are further possible sources of energy--momentum which might lead to new teleparallel solutions in $F(T)$-type and some extensions. We can also add recent papers on static radial-dependent, time-dependent, and cosmological teleparallel $F(T)$ and $F(T,B)$ type solutions suitable for universe and astrophysical models~\cite{nonvacSSpaper,staticscalarfieldSS,nonvacKSpaper,scalarfieldKS,scalarfieldTRW,coleylandrygholami,FTBcosmogholamilandry,roberthudsonSSpaper}.

However, for~electromagnetic teleparallel solutions, there are a limited number of recent contributions. There are some interesting papers on magnetic teleparallel BH solutions, especially from G.G.L. Nashed~\cite{nashedmagBH1,nashedmagBH2,nashedBH3,nashedBH4,teleBHHAL,teleBHsindian}. These papers focus essentially on typical BH solutions and usual electromagnetic situations by solving in TEGR and some preliminary cases of teleparallel gravity at the astrophysical scale. Therefore, there are no really direct papers using the TRW-based frame approach leading to electromagnetic-based cosmological teleparallel solutions using the coframe/spin-connection pair approach. This last missing piece is the key point justifying new development in this way in cosmological teleparallel $F(T)$-type gravity. {By this, we mean we will become able to study cosmological systems with electromagnetism in the teleparallel gravity framework and apply new solutions to some concrete physical problems.}

Ultimately we want to study in detail the electromagnetic TRW cosmological solutions with the physical impacts. Therefore, at the current stage, we need to find the possible electromagnetic source-based teleparallel $F(T)$-gravity solutions in a Robertson--Walker spacetime (TRW). We had found the TRW geometry and solved the TRW FEs and conservation laws (CL) for PF and SF solutions in teleparallel $F(T)$ and $F(T,B)$ gravities~\cite{coleylandrygholami,preprint,FTBcosmogholamilandry,scalarfieldTRW,ChaplyginpolyTRW}. But~we can do further and aim to solve for electromagnetic teleparallel $F(T)$ solutions as the most suitable next step of development. {We ultimately want to study the impacts of electromagnetic sources in the universe on its evolution and determine the contributions of electromagnetic sources to the universe's expansion for flat and non-flat scenarios. For~example, we will also study a special case of cosmological plasma-based electric field source.} We will use the same TRW geometry and FEs as defined in Sections~\ref{sect21} and \ref{sect22}, adapt the CLs for electromagnetic field (EMF) in Section~\ref{sect23}, and~then find the new teleparallel $F(T)$ solutions and graphical comparisons in Section~\ref{sect3}. We will then discuss the impacts of new teleparallel solutions in terms of electromagnetic fields in Section~\ref{sect51}, and~then make guidelines for experimental data-based comparison studies in Section~\ref{sect52} before concluding in Section~\ref{sect6}.

	\newpage

\section{Summary of Teleparallel Gravity and Field~Equations}\label{sect2}
\unskip

\subsection{Teleparallel $F(T)$-Gravity Theory Field Equations and Torsional~Quantities}\label{sect21}

\noindent {The} 
 {{teleparallel $F(T)$-gravity action integral with gravitational sources} is written as follows~\cite{Aldrovandi_Pereira2013,Bahamonde:2021gfp,Krssak:2018ywd,Coley:2019zld,SSpaper,nonvacSSpaper,nonvacKSpaper,roberthudsonSSpaper,scalarfieldKS,scalarfieldTRW,staticscalarfieldSS,ChaplyginpolyTRW}:}
\begin{eqnarray}\label{1000}
	S_{F(T)} = \int\,d^4\,x\,\left[\frac{h}{2\kappa}\,F(T)+\mathcal{L}_{Source}\right],
\end{eqnarray}
where $h$ is the coframe determinant, $\kappa$ is the coupling constant, and $\mathcal{L}_{Source}$ is the gravitational source terms. {The first term of Equation \eqref{1000} containing the $F(T)$ function determines the exact spacetime geometry structure type of the physical system where a gravitational source is located and evolving.} We will apply the least-action principle on Equation \eqref{1000} to find the symmetric and antisymmetric parts of FEs as follows~\cite{SSpaper,nonvacSSpaper,nonvacKSpaper,roberthudsonSSpaper,scalarfieldKS,scalarfieldTRW,staticscalarfieldSS,ChaplyginpolyTRW}:
\begin{eqnarray}
	\kappa\,\Theta_{\left(ab\right)} &=& F_T\left(T\right) \overset{\ \circ}{G}_{ab}+F_{TT}\left(T\right)\,S_{\left(ab\right)}^{\;\;\;\mu}\,\partial_{\mu} T+\frac{g_{ab}}{2}\,\left[F\left(T\right)-T\,F_T\left(T\right)\right],  \label{1001a}
	\\ 
	0 &=& F_{TT}\left(T\right)\,S_{\left[ab\right]}^{\;\;\;\mu}\,\partial_{\mu} T, \label{1001b}
\end{eqnarray}
with $\overset{\ \circ}{G}_{ab}$ as the Einstein tensor, $\Theta_{\left(ab\right)}$ the energy--momentum, $g_{ab}$ the gauge metric and $\kappa$ the coupling constant. The~torsion tensor $T^a_{~~\mu\nu}$, the~torsion scalar $T$ and the super-potential $S_a^{~~\mu\nu}$ are defined as follows~\cite{Coley:2019zld}:
\begin{eqnarray}
	T^a_{~~\mu\nu} &=& \partial_{\mu}\,h^a_{~~\nu}-\partial_{\nu}\,h^a_{~~\mu}+\omega^a_{~~b\mu}h^b_{~~\nu}-\omega^a_{~~b\nu}h^b_{~~\mu}, \label{torsionten}
	\\
	S_a^{~~\mu\nu} &=& \frac{1}{2}\,\left(T_a^{~~\mu\nu}+T^{\nu\mu}_{~~a}-T^{\mu\nu}_{~~a}\right)-h_a^{~~\nu}\,T^{\lambda\mu}_{~~\lambda}+h_a^{~~\mu}\,T^{\lambda\nu}_{~~\lambda},
	\\
	T &=&\frac{1}{2}\,T^a_{~~\mu\nu}S_a^{~~\mu\nu}.
\end{eqnarray}	
{Equation} 
 \eqref{torsionten} can be expressed in terms of the three irreducible parts of torsion tensor as follows:
\begin{eqnarray}
	T_{abc} = \frac{2}{3}\left(t_{abc}-t_{acb}\right)-\frac{1}{3}\left(g_{ab}V_c-g_{ac}V_b\right)+\epsilon_{abcd}A^d
\end{eqnarray}
where,
\begin{eqnarray}
	V_a= T^b_{~ba} ,
	\quad
	A^a=\frac{1}{6}\epsilon^{abcd}T_{bcd} ,
	\quad
	t_{abc}= \frac{1}{2}\left(T_{abc}+T_{bac}\right)-\frac{1}{6}\left(g_{ca}V_b+g_{cb}V_a\right)+\frac{1}{3}{g_{ab}}V_c.
\end{eqnarray}
We usually solve for teleparallel $F(T)$-gravity in Equations \eqref{1001a} and \eqref{1001b}. Therefore, in refs.~\mbox{\cite{preprint,coleylandrygholami,scalarfieldTRW}}, we showed that Equation \eqref{1001b} is trivially satisfied despite a non-zero spin-connection, because~the teleparallel geometry is purely symmetric. Only Equation \eqref{1001a} is non-trivial and will be explicitly solved in~details.

\subsection{Teleparallel Robertson--Walker Spacetime~Geometry}\label{sect22}

Any frame-based geometry in teleparallel gravity on a frame bundle is defined by a {\mbox{${\bf h}^a$--$\omega^a_{~bc}$}} pair and a field ${\bf X}${, and~then} must satisfy the Lie Derivative-based equations~\mbox{\cite{Coley:2019zld,MCH,preprint,coleylandrygholami,scalarfieldTRW,ChaplyginpolyTRW}:}
\begin{equation}
	\mathcal{L}_{{\bf X}}\,{\bf h}_a = \lambda_a^{~b} \,{\bf h}_b \mbox{ and } \mathcal{L}_{{\bf X}}\, \omega^a_{~bc} = 0, \label{Intro:FS2}
\end{equation}
where $\omega^a_{~bc}$ is {defined in terms of the differential coframe ${\bf h}_a$ and $\lambda_a^{~b}$ is the linear isotropy group component. {For the teleparallel $F(T)$-gravity case, we must satisfy the $R^a_{~abc}=0$ condition. For~TRW spacetime geometries on an orthonormal frame, the~${\bf h}^a$--$\omega^a_{~bc}$} pair} solutions are as follows~\cite{preprint,coleylandrygholami,scalarfieldTRW,ChaplyginpolyTRW}:
\begin{align}
	{\bf h}^a_{\;\;\mu} =& Diag\left[1, a(t)\,\left(1-k\,r^2\right)^{-1/2},\,a(t)\,r,\, a(t)\,r\,\sin\theta\right], \label{TRWcoframe}
	\\
	\omega_{122} =& \omega_{133} = \omega_{144} =  W_1(t), \quad  \omega_{234} = -\omega_{243} = \omega_{342} = W_2(t), 
    \nonumber\\ 
    \omega_{233} =& \omega_{244} = - \frac{\sqrt{1-kr^2}}{a(t)r}, \quad
	\omega_{344} =  \frac{\cot(\theta)}{a(t) r}, \label{Con:FLRW} 
\end{align}
where $W_1$ and $W_2$ {are defined in terms of $k$-parameters}:
\begin{enumerate}
	\item $k=0$: $W_1=W_2=0$,
	\item $k=+1$: $W_1=0$ and $W_2(t)=\pm\,\frac{\sqrt{k}}{a(t)}$,	
	\item $k=-1$: $W_1(t)=\pm\,\frac{\sqrt{-k}}{a(t)}$ and $W_2=0$.		
\end{enumerate}

{For} 
 any $W_1$ and $W_2$, we will obtain the same symmetric FEs set to solve {for each $k$-parameter subcases}. The~Equations \eqref{TRWcoframe} and \eqref{Con:FLRW} were found by solving Equation \eqref{Intro:FS2} and the \mbox{$R^a_{~b\mu\nu}=0$} condition as defined in ref.~\cite{Coley:2019zld}. These solutions were used in several TRW spacetime-based works~\cite{preprint,coleylandrygholami,FTBcosmogholamilandry,scalarfieldTRW,ChaplyginpolyTRW}. This spacetime structure is still explainable by a $G_6$ Lie algebra group. {From Equations \eqref{TRWcoframe} and \eqref{Con:FLRW} spin-connection pair and previously defined irreducible torsion tensor parts, we can find that the torsion scalar construction is defined in terms of $W_1$ and $W_2$ as follows (see refs.~\cite{preprint,coleylandrygholami} for detailed derivations):
\begin{eqnarray}
	T(t)=6\left(H+W_1+W_2\right)\left(H+W_1-W_2\right) ,
\end{eqnarray}
where $H=\frac{\dot{a}}{a}$ is the Hubble parameter. The~torsion scalar $T(t)$ and the TRW FEs are defined for each $k$-parameter cases, leading to new additional teleparallel $F(T)$ solutions. The~FEs are purely symmetric and valid on proper frames as shown in Equations \eqref{1001a} and \eqref{1001b} and} in refs.~\cite{preprint,coleylandrygholami,scalarfieldTRW,ChaplyginpolyTRW}. The~Equation \eqref{1001b} is trivially satisfied, and we will solve Equation \eqref{1001a} for each $k$-parameter~case.

\subsection{Einstein--Maxwell Conservation Law Solutions and Energy~Conditions}\label{sect23}

{The energy--momentum and the GR CLs are derived from the $\mathcal{L}_{Source}$ term} of Equation \eqref{1000} as follows~\cite{Aldrovandi_Pereira2013,Bahamonde:2021gfp}:
\begin{align}
	\Theta_a^{\;\;\mu}=&\frac{1}{h} \frac{\delta \mathcal{L}_{Source}}{\delta h^a_{\;\;\mu}}, \quad
	\Rightarrow\quad \overset{\ \circ}{\nabla}_{\nu}\left(\Theta^{\mu\nu}\right)=0 , \label{1001e}
\end{align}
where $\overset{\ \circ}{\nabla}_{\nu}$ the covariant derivative and $\Theta^{\mu\nu}$ the conserved energy--momentum tensor. The~antisymmetric and symmetric parts of $\Theta_{ab}$ are written as follows~\cite{SSpaper,nonvacSSpaper,nonvacKSpaper,roberthudsonSSpaper,scalarfieldKS,scalarfieldTRW,staticscalarfieldSS}:
\begin{equation}\label{1001c}
	\Theta_{[ab]}=0,\qquad \Theta_{(ab)}= T_{ab},
\end{equation}
where $T_{ab}$ is the symmetric part of $\Theta^{\mu\nu}$. {Equation \eqref{1001e} forces} the symmetry of $\Theta^{\mu\nu}$ and then Equation \eqref{1001c} condition. Equation \eqref{1001c} is valid {for a matter field interacting with a metric $g_{\mu\nu}$ defined by the coframe ${\bf h}^a_{\;\;\mu}$ and the gauge $g_{ab}$,} and is not directly coupled to the $F(T)$-gravity {function}. This consideration is only valid for the null hypermomentum case (i.e., $\mathfrak{T}^{\mu\nu}=0$) as discussed in refs.~\cite{golov3,nonvacSSpaper,nonvacKSpaper,roberthudsonSSpaper,scalarfieldKS,scalarfieldTRW,staticscalarfieldSS}. This condition is defined from \mbox{Equations \eqref{1001a} and \eqref{1001b}} as follows~\cite{golov3}:
\begin{align}\label{1001h}
	\mathfrak{T}_{ab}=\kappa\Theta_{ab}-F_T\left(T\right) \overset{\ \circ}{G}_{ab}-F_{TT}\left(T\right)\,S_{ab}^{\;\;\;\mu}\,\partial_{\mu} T-\frac{g_{ab}}{2}\,\left[F\left(T\right)-T\,F_T\left(T\right)\right]=0.
\end{align}
There are more general teleparallel $\mathfrak{T}^{\mu\nu}$ definitions and $\mathfrak{T}^{\mu\nu}\neq 0$ CLs, but~{it does not affect the current} teleparallel $F(T)$-gravity {situation} 
 \cite{hypermomentum1,hypermomentum2,hypermomentum3,golov3}.

For a TRW spacetime geometry defined by Equations \eqref{TRWcoframe} and \eqref{Con:FLRW}, Equation \eqref{1001e} for $\rho=\rho_{em}$, $P_r=P_{em\,r}$ and $P_t=P_{em\,t}$ fluid equivalent for electromagnetic source is as follows~\cite{preprint,coleylandrygholami,scalarfieldTRW,roberthudsonSSpaper}:
\begin{align}\label{1003}
	\dot{\rho}_{em}+H\,\left(3\rho_{em}+P_{em\,r}+2P_{em\,t}\right)=0, \quad \text{and}\quad 2\left(P_{em\,r}-P_{em\,t}\right)+r\,\partial_r\,P_{em\,r}=0.
\end{align}
The Einstein--Maxwell Lagrangian and then the energy--momentum tensor are defined as follows~\cite{electro1,electro2,electro3,electro4}:
\begin{align}\label{1004}
	\mathcal{L}_{source}=-\frac{1}{4}\,F_{\mu\nu}F^{\mu\nu}\quad\Rightarrow\quad\Theta_{\mu\nu}=& F_{\mu\alpha}F^{\alpha}_{\;\;\nu}-\frac{1}{4}g_{\mu\nu}F^2 ,
\end{align}
where $F_{\mu\nu}=\nabla_{\mu}A_{\nu}-\nabla_{\nu}A_{\mu}$ is the electromagnetic tensor defined in terms of quadripotential $A_{\mu}$. In~terms of electric $\vec{E}$ and magnetic $\vec{B}$ fields, Equation \eqref{1004} is defined as follows:
\begin{align}\label{1004a}
	\Theta_{\mu\nu} =&\left[\begin{array}{cc}
		\frac{E^2+B^2}{2}	& \left[\vec{S}\right]^T \\
		\left[\vec{S}\right]	&  -\left[\sigma_{ij}\right]
	\end{array}\right] , 	 
\end{align}
where $\vec{S}=\vec{E}\,\times\,\vec{B}$ is the Poynting vector and $\sigma_{ij}=\left(E_iE_j-\frac{\delta_{ij}}{2}E^2\right)+\left(B_iB_j-\frac{\delta_{ij}}{2}B^2\right)$. The~Equation \eqref{1004a} is diagonalizable and can be expressed in terms of density--pressure equivalent expressions. The~diagonal form is written as follows:
\begin{align}
	\Theta^a_{\;\;\mu} \equiv Diag\left[\rho_{em},\,P_{em\,r},\,P_{em\,t},\,P_{em\,t}\right] .
\end{align}
For any EoS and/or equivalent relationship, there are energy conditions (ECs) to satisfy for any physical system based on a PF~\cite{Kontou:2020bta}:
\begin{itemize}
	\item Weak Energy Condition (\textbf{{WEC}}): 
 $\rho \geq 0$, $P_r+\rho \geq 0$ and $P_t+\rho \geq 0$.
	
	\item Strong Energy Condition (\textbf{{SEC}}): $P_r+2P_t+\rho \geq 0$, $P_r+\rho \geq 0$ and $P_t+\rho \geq 0$.

	\item Null Energy Condition (\textbf{{NEC}}): $P_r+\rho \geq 0$ and $P_t+\rho \geq 0$. 
	
	\item Dominant Energy Condition (\textbf{{DEC}}): $\rho \geq |P_r|$ and $\rho \geq |P_t|$.
\end{itemize}

{By} 
 this way, we will verify the physical consistency of all CL solutions found in the current~paper.

There are three main~cases:
\begin{enumerate}
	\item \textbf{{General electromagnetic universe:}} For any $\vec{E}\neq \vec{0}$ and $\vec{B} \neq 0$, Equation \eqref{1004a} becomes the following:
\begin{align}\label{1030a}
		\Theta_{\mu\nu} =&\left[\begin{array}{cccc}
			\frac{E^2+B^2}{2}	&  S_1 &  S_2 & S_3 \\
			S_1 &  -\sigma_{11} & -\sigma_{12} & -\sigma_{13} \\
			S_2 &  -\sigma_{12} & -\sigma_{22} & -\sigma_{23} \\
			S_3 &  -\sigma_{13} & -\sigma_{23} & -\sigma_{33} \\
		\end{array}\right] . 	 
	\end{align}
	By setting $E_2=E_3=E_t$ and $B_2=B_3=B_t$, we find that $S_1=S_r=0$, $S_2=-S_3=S_t=0$ leading to $E_r\,B_t=E_t\,B_r$ for consistency. By~using the last constraint and then by diagonalisation, we find that $P_{em\,t}=-P_{em\,r}=\rho_{em}=\frac{E^2}{2}\left(1+\frac{B_r^2}{E_r^2}\right)$ and the \textbf{{WEC}}, \textbf{{SEC}}, \textbf{{NEC}} and \textbf{{DEC}} are all satisfied by the $E^2 \geq 0$, $E_r^2\geq 0$ and $B_r^2 \geq 0$ conditions {for any teleparallel $F(T)$ solution}. Then Equation \eqref{1003} becomes the following:
\begin{align}\label{1030b}
		\dot{\rho}_{em}+4H\,\rho_{em}=0, \quad \text{and}\quad 4\rho_{em}+r\,\partial_r\,\rho_{em}=0,
	\end{align}
	From the $2$nd CL, we will find that $\rho_{em}(t,r)=\frac{\rho_{em}(t,0)}{r^4}$. Then the $1$st CL solution in terms of torsion scalar $T$ is exactly the following:
\begin{align}\label{1030c}
		\rho_{em}(t(T),r(T))=\rho_{em}(T)=\frac{\rho_{em}(0)}{r^4(T)\,a^{4}(T)}.
	\end{align}

	\item \textbf{{Pure electric universe} $|\vec{B}| \ll |\vec{E}|$ limit:} Equation \eqref{1004a} becomes the following:
\begin{align}\label{1010a}
		\Theta_{\mu\nu} =&\left[\begin{array}{cc}
			\frac{E^2}{2}	& 0 \\
			0	&  \left[\frac{\delta_{ij}}{2}E^2-E_iE_j\right]
		\end{array}\right] .	 
	\end{align}
	By diagonalization techniques applied on Equation \eqref{1010a} and setting $E_2=E_3=E_t$, we find that $P_{em\,t}=-P_{em\,r}=\rho_{em}=\frac{E^2}{2}$, and~then the ECs are trivially satisfied by the $E^2 \geq 0$ condition {for any teleparallel $F(T)$ solution}. The~Equations \eqref{1030b} and \eqref{1030c} will still be applicable for a pure electric~universe.

	\item \textbf{{Pure magnetic universe} $|\vec{B}| \gg |\vec{E}|$ limit:} Equation \eqref{1004a} becomes the following:
\begin{align}\label{1020a}
		\Theta_{\mu\nu} =&\left[\begin{array}{cc}
			\frac{B^2}{2}	& 0\\
			0	&  \left[\frac{\delta_{ij}}{2}B^2-B_iB_j\right]
		\end{array}\right] . 	 
	\end{align}
	Still by using diagonalization techniques on Equation \eqref{1020a} and setting $B_2=B_3$, we find that $P_{em\,t}=-P_{em\,r}=\rho_{em}=\frac{B^2}{2}$, and~then the ECs are trivially satisfied by the $B^2 \geq 0$ condition {for any teleparallel $F(T)$ solution}. The~Equations \eqref{1030b} and \eqref{1030c} will still be applicable for a pure magnetic~universe.

\end{enumerate}


\section{Electromagnetic Teleparallel Field Equations~Solutions}\label{sect3}

The general FEs system for TRW cosmological spacetimes are written as follows~\cite{preprint,coleylandrygholami,scalarfieldTRW,ChaplyginpolyTRW}:
\begin{enumerate}
	\item {${\bf k=0}$} \textbf{{flat or non-curved}}:
\begin{align}
		\kappa\rho_{em}	= &	-\frac{F}{2}+6H^2\,F_T ,  \label{1010}
		\\
		\kappa(\rho_{em}+P_{em,\,r}+2P_{em,\,t}) =&	F-6\left( \dot{H}+2H^2\right)F_T-6H\,F_{TT}\dot{T}, \label{1011}
		\\
		T =& 6H^2  .\label{1012}
	\end{align}
	The Equation \eqref{1012} yields to $H=\sqrt{\frac{T}{6}}$ and from Section~\ref{sect23} results, we will find that $\kappa(\rho_{em}+P_{em,\,r}+2P_{em,\,t})=2\kappa\,\rho_{em}$. In~this case, Equations \eqref{1010} and \eqref{1011} become the following:
\begin{align}
		2\kappa\rho_{em} = & -F+2TF_T	, \label{1013a}
		\\	
		2\kappa\,\rho_{em} =& F-\left(6 \dot{H}+2T\right)F_T-\sqrt{6T}\,F_{TT}\dot{T}. \label{1013b}
	\end{align}
	By merging Equations \eqref{1013a} and \eqref{1013b}, we find the unified FE, written as follows:
\begin{align}
		0 = F-\left( 3\dot{H}+2T\right)F_T-\sqrt{\frac{3}{2}\,T}\,F_{TT}\dot{T}.  \label{1014}
	\end{align}
	The pure vacuum solution ($\rho_{em}=0$) to Equation \eqref{1013a} is $F(T)=F_0\,\sqrt{T}$. However, for~$\rho_{em} \neq 0$, we can set $a(t)=a_0\,t^n$ as cosmological scale and consider the radial coordinate $r=r(T)$ as a complementary function allowing $P_{em,\,r}\neq P_{em,\,t}$ situations. Equation \eqref{1014} and solutions are written as follows:
\begin{align}
		&	0= nF+\left(\frac{1-4n}{2}\right)TF_T+T^{2}\,F_{TT}, &  \label{1015}
		\\
		\Rightarrow\,	F(T)=& F_0\,\sqrt{T}+F_1\,T^{2n} , \quad  & n\neq \frac{1}{4}, \label{1017}
		\\
		=& F_0\,\sqrt{T}+F_1\,\sqrt{T}\,\ln(T) , \quad  & n = \frac{1}{4}. \label{1018}
	\end{align}
	Equations \eqref{1017} and \eqref{1018} are double power-law teleparallel solution very similar to those found in some recent teleparallel Robertson--Walker based papers~\cite{coleylandrygholami,scalarfieldTRW,FTBcosmogholamilandry,ChaplyginpolyTRW}.
	
	\item {${\bf k=-1}$} \textbf{{negative curved}}:
\vspace{-9pt} 
\begin{align}
		\kappa\rho_{em}(T)	= & -\frac{F}{2}+6\,H\left(H+\frac{\delta\sqrt{-k}}{a}\right) F_T, \label{1020}
		\\
		\kappa(\rho_{em}+P_{em,\,r}+2P_{em,\,t}) =&	F-6\left(\dot{H}+H^2+\left(H+\frac{\delta\sqrt{-k}}{a}\right)^2 \right) F_T -6\left(H+\frac{\delta\sqrt{-k}}{a}\right)\,F_{TT}\dot{T}, \label{1021}
		\\
		T =& 6\left( H+ \frac{\delta\sqrt{-k}}{a}\right)^2  . \label{1022}
	\end{align}
    \normalsize
	From Equation \eqref{1022} and using $a(t)=a_0\,t^n$ ansatz, we find a characteristic equation yielding to $t(T)$ solutions, written as follows:
\begin{align}
		& 0= \frac{\delta\sqrt{-k}}{a_0}\,t^{-n}+n t^{-1}-\delta_1\sqrt{\frac{T}{6}}. \label{1023}
	\end{align}
	By substitution of the $\kappa(\rho_{em}+P_{em,\,r}+2P_{em,\,t})=2\kappa\,\rho_{em}$ relation and merging Equations \eqref{1020} and \eqref{1021}, we find the unified FE, written as follows:
\begin{align}
		0 =&	F-3\left(\frac{n(n-1)}{t^2(T)}+ \frac{2n\delta_1}{t(T)}\sqrt{\frac{T}{6}} +\frac{T}{6} \right) F_T +T\, \left(\frac{n}{t^2(T)}+\frac{n\delta\sqrt{-k}}{a_0}\,t^{-n-1}(T)\right)\,F_{TT}. \label{1041}
	\end{align}
	The possible solutions for Equation \eqref{1023} are by using the far-future approximation, written as follows ($t(T)\gg 1$ as in ref.~\cite{ChaplyginpolyTRW}, except~for $n=1$ subcase):
	\begin{enumerate}
		\item {${\bf n=\frac{1}{2}}$} (slow expansion and $+$ solution):
\begin{align}
			& t^{-1}(T)=\left[-\frac{\delta\sqrt{-k}}{a_0}\pm \sqrt{-\frac{k}{a_0^2}+\delta_1\sqrt{\frac{2T}{3}}}\right]^2 \approx  \frac{a_0^2}{6(-k)}\,T . \label{1023a}
		\end{align}
		By substitution, Equation \eqref{1041} is found by neglecting $\frac{a_0^4}{24(-k)^2}\,T$ terms, written as follows:
        \small
\begin{align}\label{1041a}
			0 \approx &	 F+\frac{1}{2}\left(\frac{A^4}{4}\,T- A\,T^{1/2} -1 \right) TF_T + \frac{A}{12} \left(A\,T^{1/2}+1\right)\,T^{5/2}\,F_{TT},
			\nonumber\\	
			\Rightarrow\;F(T)=& \frac{1}{{\sqrt{T}}\, \left(1+A \sqrt{T}\right)^{2}} \Bigg[F_{1} \left(3 A^{3} T^{\frac{3}{2}}+10 A^{2} T+11 A \sqrt{T}+4\right) \exp\left({-\frac{12}{A \sqrt{T}}}\right)
			\nonumber\\
			&\quad+F_{2} \Bigg(\exp\left({-\frac{12 \left(1+A \sqrt{T}\right)}{A \sqrt{T}}}\right) \left(A^{3} T^{\frac{3}{2}}+\frac{10 A^{2} T}{3}+\frac{11 A \sqrt{T}}{3}+\frac{4}{3}\right)
           \nonumber\\
			&\quad \times\, \text{Ei}_{1}\! \left(-\frac{12 \left(1+A \sqrt{T}\right)}{A \sqrt{T}}\right)
			+\frac{119 A^{3} T^{\frac{3}{2}}}{1296}+\frac{11 A^{2} T}{54}+\frac{A \sqrt{T}}{9}\Bigg)\Bigg] ,	 
		\end{align}
        \normalsize
		where $A=\frac{\delta_1\,a_0^2}{\sqrt{6}(-k)}$. For~the very-far-future approximation: Equation \eqref{1041a} becomes $0\approx F-\frac{T}{2}F_T$ and the solution is $F(T)\,\rightarrow\, F_1\,T^2$.
		
		\item {${\bf n=1}$} (linear expansion):
\begin{align}
			&  t^{-1}(T)= \frac{\delta_1}{\left(\frac{\delta\sqrt{-k}}{a_0}+1\right)}\sqrt{\frac{T}{6}}. \label{1023b}
		\end{align}
		By substitution, Equation \eqref{1041} becomes the following:
\begin{align}\label{1041b}
			0 =&  F-\left(\frac{1}{\left(\frac{\delta\sqrt{-k}}{a_0}+1\right)} +\frac{1}{2} \right) TF_T + \frac{1}{6\left(\frac{\delta\sqrt{-k}}{a_0}+1\right)}\,T^2F_{TT}, 
            \nonumber\\
            \quad
			&\Rightarrow\;F(T)= F_1\,T^{r_{+}}+F_2\,T^{r_{-}} ,
		\end{align}
		where $r_{\pm}= 5+\frac{3\delta\sqrt{-k}}{2a_0}\pm \sqrt{19+\frac{9\delta\sqrt{-k}}{a_0}-\frac{9k}{4a_0^2}}$.
		
		\item {${\bf n=2}$} (fast expansion and $+$ solution):
\begin{align}
			& t^{-1}(T)=-\frac{\delta a_0}{\sqrt{-k}} \pm \sqrt{-\frac{a_0^2}{k}+\delta_1 \delta a_0\sqrt{-\frac{T}{6k}}} \approx \delta_1 \sqrt{\frac{T}{24}} 
			. \label{1023c}
		\end{align}
		By substitution, Equation \eqref{1041} becomes the following:
\begin{align}\label{1041c}
			0 \approx &	 F-\frac{7}{4}\,TF_T +\frac{1}{12}\,T^2F_{TT},\quad \Rightarrow\;F(T)= F_1\,T^{11+\sqrt{109}}+F_2\,T^{11-\sqrt{109}}.	
		\end{align}
		
		\item {${\bf n\,\rightarrow\,\infty}$} (very fast expansion limit):
\begin{align}
			& t^{-1}(T)\approx \frac{\delta_1}{n}\sqrt{\frac{T}{6}}\,\rightarrow\,0. \label{1023d}
		\end{align}
		By substitution, Equation \eqref{1041} becomes $F=2TF_T$ and then $F(T)=F_0\,\sqrt{T}$. The~Equation \eqref{1020} leads to $\rho_{em}(T)\approx \frac{F_0\,\left(\delta_1-1\right)}{2}\sqrt{T}=0$ for a positive \mbox{Equation~\eqref{1023d}} (i.e., $\delta_1=1$), the~cosmological vacuum solution, where no EM~subsists.

	\end{enumerate}

	\item {${\bf k=+1}$} \textbf{{positive curved}}:
\begin{align}
		\kappa\rho_{em}(T)	= &	-\frac{F}{2}+6H^2\,F_T ,  \label{1030}
		\\
		\kappa(\rho_{em}+P_{em,\,r}+2P_{em,\,t}) =&	F-6\left( \dot{H}+2H^2-\frac{k}{a^2}
		\right)F_T-6H\,F_{TT}\dot{T}, \label{1031}
		\\
		T =& 6\left[ H^2- \frac{k}{a^2}\right]  . \label{1032}
	\end{align}
	From Equation \eqref{1032} and using $a(t)=a_0\,t^n$ ansatz, we find the characteristic equation for $t^{-1}(T)$:
\begin{align}
		&\frac{k}{a_0^2}t^{-2n}-n^2\,t^{-2}+\frac{T}{6}=0 . \label{1033}
	\end{align}
	We simplify and unify by substitution of $\kappa(\rho_{em}+P_{em,\,r}+2P_{em,\,t})=2\kappa\,\rho_{em}$ the \mbox{Equations \eqref{1030} and \eqref{1031}}:
\vspace{-9pt}
\begin{align}
		0  =& 2F-\left( 6n(3n-1)\,t^{-2}(T)+T\right)F_T+12n^2\,t^{-2}(T)\left(T-6n(n-1)t^{-2}(T)\right)\,F_{TT} . \label{1051}
	\end{align}
	The possible solutions of Equation \eqref{1033} are with the far-future approximation ($t(T)\gg 1$ as in ref.~\cite{ChaplyginpolyTRW}, except~for $n=1$ subcase):
	\begin{enumerate}
		\item {${\bf n=\frac{1}{2}}$} (slow expansion and $-$ solution):
\begin{align}
			&t^{-1}(T)= \frac{2k}{a_0^2}\pm\sqrt{\frac{4k^2}{a_0^4}+\frac{2T}{3}} \approx -\frac{a_0^2}{6k}\,T . \label{1033a}
		\end{align}
		By substitution, Equation \eqref{1051} becomes the following:
		\small
\begin{align}\label{1051a}
			0 \approx &  2F-\left(1+\frac{a_0^4}{24k^2}\,T\right)TF_T+\frac{a_0^4}{12k^2}\left(1+\frac{a_0^4}{24k^2}\,T\right)\,T^3F_{TT}     ,
			\nonumber\\	
			\Rightarrow\;F(T)=& \exp\left(-\frac{12k^2}{a_0^4\,T}\right)\,\mathrm{HeunC}\! \left(\frac{1}{2},1,-1,-\frac{1}{2},0,-\frac{24k^2}{a_0^4\,T}\right) 
            \nonumber\\
            & \quad\times\,\Bigg[ F_{1}\,+F_{2}\,\int\,\frac{\exp\left(\frac{12k^2}{a_0^4\,T}\right)\,dT}{\mathrm{HeunC}\! \left(\frac{1}{2},1,-1,-\frac{1}{2},0,-\frac{24k^2}{a_0^4\, T}\right)^{2}}\Bigg].
		\end{align}
		\normalsize
		For the very-far-future approximation: Equation \eqref{1051a} becomes $0\approx F-\frac{T}{2}F_T$ leading to $F(T)\,\rightarrow\, F_1\,T^2$ as for the $k=-1$ case.
		
		\item {${\bf n=1}$} (linear expansion):
\begin{align}
			t^{-2}(T)=\frac{T}{6\left(1-\frac{k}{a_0^2}\right)}. \label{1033b}
		\end{align}
		By substitution, Equation \eqref{1051} becomes the following:
\begin{align}\label{1051b}
			0 =&  2F-\left(\frac{2}{\left(1-\frac{k}{a_0^2}\right)}+1\right)TF_T+\frac{2}{\left(1-\frac{k}{a_0^2}\right)}\,T^2F_{TT} ,
            \nonumber\\
            \quad
			&\Rightarrow\;F(T)= F_1\,T^2+F_2\,T^{\frac{1}{2}\left(1-\frac{k}{a_0^2}\right)} .
		\end{align}

		\item {${\bf n=2}$} (fast expansion and $-$ solution):
\begin{align}
			& t^{-2}(T)= \frac{2a_0^2}{k} \pm\sqrt{\frac{4a_0^4}{k^2}-\frac{a_0^2\,T}{6k}} \approx \frac{T}{24}	. \label{1033c}
		\end{align}
		By substitution, Equation \eqref{1051} becomes the following:
\begin{align}\label{1051c}
			0 \approx &\,  2F-\frac{7}{2}\,TF_T+T^2F_{TT} ,\quad \Rightarrow\;F(T)= F_0\sqrt{T}+F_1\,T^4.	
		\end{align}

		\item {${\bf n\,\rightarrow\,\infty}$} (very fast expansion limit):
\begin{align}
			t^{-2}(T)\,\approx \frac{T}{6n^2}\,\rightarrow\,0. \label{1033d}
		\end{align}		
		By substitution, Equation \eqref{1051} becomes $F=2TF_T$, leading to $F(T)=F_0\,\sqrt{T}$ and then Equation \eqref{1030} becomes $\kappa\rho_{em}(T)=0$ as for the $k=-1$ case, a~pure electromagnetic~vacuum.

	\end{enumerate}

\end{enumerate}

\begin{figure}[ht]
	\hspace{-15pt}\includegraphics[width=6.5cm,height=6.5cm]{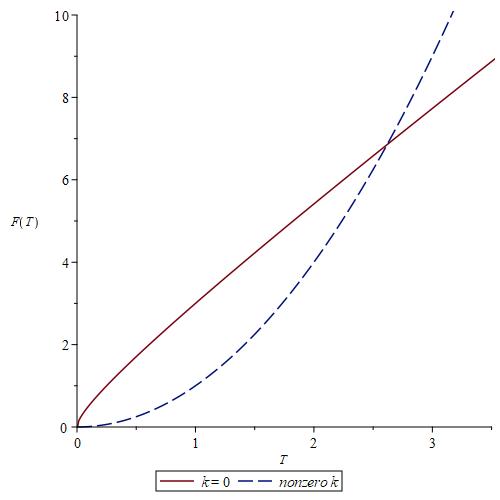} 
	\includegraphics[width=6.5cm,height=6.5cm]{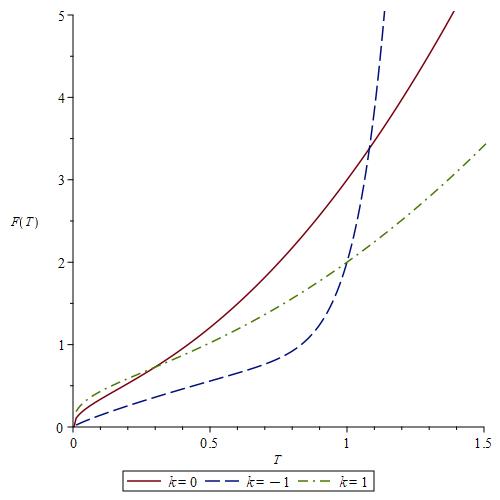}   
	\centering{\includegraphics[width=6.5cm,height=6.5cm]{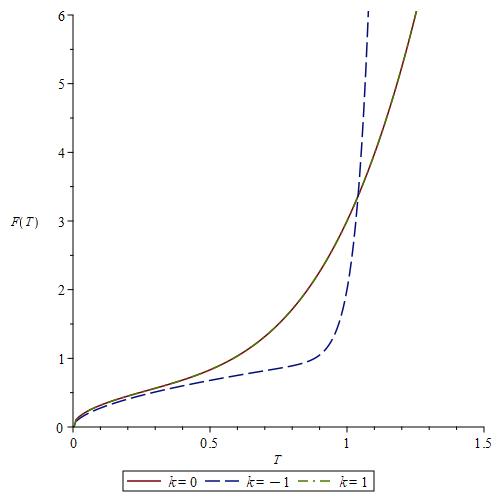}  }
	\caption{Plot of teleparallel $F(T)$ solutions for $k=0$ and $\pm 1$ (\textbf{left}: $n=\frac{1}{2}$, \textbf{right}: $n=1$, and~\textbf{bottom}: $n=2$. Curves are for $F_1=F_2=2F_0$ setting).}
	\label{figure1}
\end{figure}

In Figure~\ref{figure1}, we compare the new teleparallel $F(T)$ solutions for highlighting the main common points between {typical $k=-1,\,0\,+1$ solutions. We want to highlight the impacts of the $k$-parameter on new EMF-based teleparallel solutions}. For~$n=\frac{1}{2}$ subcase (slow universe expansion), we find that the $k=\pm 1$ cases lead to the same quadratic $F(T)$ limit for the very-far-future limit ($t(T)\gg 1$ approximation), while the $k=0$ case leads to the superposition of the TEGR-like term with the large $n$ limit $\sqrt{T}$ term. For~the $n=1$ (linear universe expansion) subcase, we find three different curves of teleparallel $F(T)$ solutions. For~$n=2$ subcase (fast universe expansion), we find that the $k=0$ and $+1$ cases have the same superposed curves and $k=-1$ is different. However, we have obtained in the current section new simple analytical EMF source-based teleparallel $F(T)$ solutions suitable for any electromagnetic-based universe models. {From the teleparallel $F(T)$ solutions, we can find the $\rho_{em}(T)$ expression and then the related $E(T)$ and/or $B(T)$ EMF solutions. We will then have to verify the Maxwell equations as stated below in Section~\ref{sect51}.}
\unskip

\newpage

\section{Physical Interpretations and Experimental Data~Comparisons}\label{sect5}
\unskip

\subsection{Electromagnetic Field~Interpretations}\label{sect51}

From the CLs solutions as found in Equation \eqref{1030c}, we can find for each new teleparallel $F(T)$ solution the corresponding EMF $E$ and/or $B$. By~using the same $a(t)=a_0\,t^n$ ansatz, the~Equation \eqref{1030c} is written as follows:
\begin{align}\label{5000}
	\rho_{em}(T)=\frac{\tilde{\rho}_{em}(0)}{\left(t^{n}(T)\,r(T)\right)^4},
\end{align} 
where $\tilde{\rho}_{em}(0)=\frac{\rho_{em}(0)}{a_0^4}$. We find from Equation \eqref{5000} that any EMF will be described by $E(T)=\frac{E_0}{\left(a(T)\,r(T)\right)^2}$, a~Coulombian field, and/or $B(T)=\frac{B_0}{\left(a(T)\,r(T)\right)^2}$, a~magnetic dipole field, with~an expanding universe term. In~the far future, we find that the electromagnetic fields will be decreasing and then becoming more negligible. But~the EMF as a CL solution also needs to satisfy the Maxwell equations as follows~\cite{electro1,electro2,electro3,electro4}:
\begin{align}\label{5001}
	\nabla_{\mu}\,F^{\mu\nu} =& J^{\nu}, \quad\text{and} \quad\nabla_{\kappa}F_{\mu\nu}+\nabla_{\nu}F_{\kappa\mu}+\nabla_{\mu}F_{\nu\kappa}=0.
\end{align}
We are in principle able to find the four-current $J^{\nu}$ and the four-potential $A^{\nu}$ averaged expressions of the universe from the first Equation \eqref{5001}, $F^{\mu\nu}$ definition, and~using the $E(T)$ and/or $B(T)$ expressions found from Equation \eqref{5000}. The~result will depend on the situation: a pure electric (or magnetic) or a general EMF of universe. In~principle, each teleparallel $F(T)$ solution of each cosmological case and subcase will lead to different $J^{\nu}$ and $A^{\nu}$ expressions. We also need to consider that $J^{\nu}$ is also a conserved current satisfying $\nabla_{\nu}J^{\nu}=0$ and to set the electromagnetic gauge for $A^{\nu}$. All for satisfying the gauge-invariance fundamental principle for electromagnetic fields~\cite{electro1,electro2,electro3,electro4}. We can do this type of development for each teleparallel solutions, $F(T)$, and also for~extensions.

{For example, we will find the corresponding EMF of the $k=0$ case teleparallel solution described by Equations \eqref{1017} and \eqref{1018}. By~using the first FE described by Equation \eqref{1010}, we find the the following:
\begin{align}
\rho_{em}(T) =& \frac{(4n-1)}{2\kappa}\,F_1\,T^{2n} , \quad  & n\neq \frac{1}{4}, \label{5002}
\\
=& \frac{F_1}{\kappa}\,\sqrt{T} , \quad  & n= \frac{1}{4}. \label{5003}
\end{align}
Equations \eqref{5002} and \eqref{5003} lead to the $\rho_{em}(T)=\rho_{em}(0)\,T^{2n}$. Using CLs of Section~\ref{sect23}, we find that $E(T)=E_0\,T^n$ and/or $B(T)=B_0\,T^n$. From~Equation \eqref{5000} and finding that $a^2(T)=\tilde{a}_0\,T^{-n}$, we will find that $r(T)=r_0=$ constant. The~current cosmological system does not depend on the radial coordinate and only depends on the time coordinate. We confirm that $T=T(t)$ and then $\vec{E}=\vec{E}(t)$ and/or $\vec{B}=\vec{B}(t)$, an~induction-based~EMF.

We can also find the potential $A^{\mu}$ and $J^{\mu}$ from Equation \eqref{5001}, but~there are several possible solutions according to the gauge invariance principle. By~using the vectorial form of Maxwell's equations, we will find that the magnetic field is $\vec{B}=\vec{0}$, and~then an induced time-dependent electric current $\vec{J}_{ind}=-\partial_t\,\vec{E}=\vec{\tilde{J}}_0\,T^{n+\frac{1}{2}}$. We find a pure electric field case solution satisfying all Maxwell equations stated by Equation \eqref{5001}.

As a good application, we can assume a plasma-based universe model where \mbox{$\vec{J}_{ind}=\sigma_{pl}\vec{E}$} leads to the Maxwell equation $\sigma_{pl}\vec{E}_{pl}=-\partial_t\,\vec{E}_{pl}$ \cite{electro1,electro2,electro3,electro4,plasmaelectric}. From~this last equation, we search for a $k=0$ cosmological solution:
\begin{align}\label{5004}
	\vec{E}_{pl}(T)=& \vec{E}_0\,\exp\left(-\sigma_{pl}\,t(T)\right)=\vec{E}_0\,\exp\left(-\frac{\sqrt{6}\,n\sigma_{pl}}{\sqrt{T}}\right),
\nonumber\\
			=& \vec{E}_0\,\sum_{n=0}^{\infty}\,\frac{\left(-\sqrt{6}\,n\sigma_{pl}\right)^n}{n!}\,T^{-n/2} .
\end{align}
Such $k=0$ flat cosmological plasma models can be seen as a large superposition of power-law term solutions. We just constrain the induced current to be proportional to the plasma electric field $\vec{E}_{pl}$. We can also proceed with the same type of development for the non-flat cosmological solutions ($k=\pm 1$ cases). We can also see in Equation \eqref{5004} that a high conductivity will lead to a weak electric field for satisfying the Maxwell~equations.

}

\subsection{Experimental Data Comparison~Guidelines}\label{sect52}

\noindent {The} 
 new electromagnetic teleparallel $F(T)$ solutions need to be compared and tested with existing experimental data sets from experiments such as Dark Energy Spectroscopic Instrument (DESI), and~other Baryonic Data (BAO) and any other cosmological redshift measurements ($H(z)$-based measurements) \cite{DESI1,DESI2,DESI3,DESI4,DESI5,BAOvsSN1,BAO2}. The~far-future teleparallel solution for $k=0$ cases can be compared to the baryonic-based cosmological background data sets. The~$k=\pm 1$ teleparallel $F(T)$ solutions for the far-future universe need to be compared with data for determining the suitable non-flat cosmological models with EMF contributions. We can now determine the most realistic solution for universe models and explanations in terms of electromagnetic averaged~contributions.

Therefore, {this paper primarily aims to find by a mathematical--physics approach the main relevant teleparallel $F(T)$ solutions for usual electromagnetic sources}. We find in Section~\ref{sect3} the most relevant and verifiable new teleparallel $F(T)$ solutions to be tested and compared with experimental data sets. {We specifically find that the $k=0$ case leads to a pure electric field source expressible as a power-law in torsion scalar, i.e.,~$|\vec{E}|\sim T^{2n}$. We have then shown that the same solution is generalizable to an exponential cosmological plasma electric field source solution.} We can propose future works aiming to compare, by data fitting, the new teleparallel $F(T)$ solutions with background measurements, as those made by the DESI collaboration~\cite{DESI1,DESI2,DESI3,DESI4,DESI5}. We have seen that a number of new teleparallel $F(T)$ solutions are close to those of refs.~\cite{myrzacosmotele1,myrzacosmospin,paliacosmo}. These results will allow us to make good comparisons with data sets in future data-fitting-based works for determining which of the new solution classes are the most realistic for the EMF contribution of universe~models.

We will be able to better confirm and/or adapt the $\Lambda$CDM models to the data sets by determining the most suitable new teleparallel $F(T)$ solution models in terms of electromagnetic contributions. Some recent similar studies using data comparison have been performed for more simple universe models using redshift $H(z)$, BAO, and other similar data sets (see refs.~\cite{celiaBAO1,celiaBAO2,celiaBAO3} and refs. within). It is possible to use the scale factor $a$ and determine the $n$-parameter possible values for the comparison with redshift measurements. Some data analysis techniques used in the mentioned works are reusable for future comparative works for the new electromagnetic source teleparallel $F(T)$ solutions found here. But~the data comparison goes beyond the aims and scopes of the theoretical and mathematical physics-based approach of the current paper. But~the data fitting based studies need to be performed in a near future, as suggested in some recent works for teleparallel $F(T)$ and $F(T,B)$ type solutions~\cite{FTBcosmogholamilandry,ChaplyginpolyTRW}. We have all the ingredients to achieve a possible future full data-based EMF contribution study of universe~models.

{Reusing the same Section~\ref{sect51} example, we should illustrate how we can compare redshift data with the flat cosmological ($k=0$ case) teleparallel $F(T)$ solution. If~$H(T)=\sqrt{\frac{T}{6}}$ and $1+z=t^{-n}(T(z))=\left(\frac{T}{6n^2}\right)^{n/2}$, we then find $H(z)=n\,\left(1+z\right)^{1/n}$. From~any redshift data $z$ measurement set and any teleparallel cosmological solution, we are able to find the $H(z)$, $T(z)$ and then data suitable teleparallel $F(T)$ solutions~\cite{telecosmoreconstruct,gravredshiftrev}. We also find as a relative density with the electric field $\Omega_{em}(z)=\frac{\rho_{em}(z)}{\rho_0}=\frac{|\vec{E}(z)|^2}{|E_0|^2}\sim T^{2n}(z)\equiv\left(1+z\right)^4$ (and/or $\frac{|\vec{B}(z)|^2}{|B_0|^2}$ for non-flat cosmology). We can do a similar development for $k=\pm 1$ cases under the same principle, but~we will find some more elaborated expressions. We can really make the link between the data sets and teleparallel solution source terms in this~manner.

}


\section{Concluding~Remarks}\label{sect6}

We will first conclude this paper by the flat cosmological $k=0$ teleparallel $F(T)$ solutions, which are described by double power-law functions. {We also find that a power-law electric field ($|\vec{E}(T)|\sim T^n$) constitutes the only possible energy--momentum source term satisfying the Maxwell equations. This last solution can be generalized by superposition to a cosmological plasma solution. We have also found that we can link any redshift $z$ data set to scale factor $a(z)$, Hubble parameter $H(z)$, torsion scalar $T(z)$, and then the new teleparallel $F(T)$ solutions in Section~\ref{sect5} developments.}

For the spatially curved cosmological $k=\pm 1$ teleparallel $F(T)$ solutions are described by various forms. For~$n=\frac{1}{2}$, the~general solutions are described by special function but~go to the same quadratic term $F(T) \sim T^2$ in the very-far-future limit for $k=\pm 1$. However, for $n=1$ and $2$ (faster universe expansion), the~double polynomial function form describes the teleparallel $F(T)$ solutions. Under~these considerations, we claim that non-flat $k=\pm 1$ teleparallel $F(T)$ solutions also have some common points with several cosmological teleparallel $F(T)$ and $F(T,B)$ solutions of refs.~\cite{coleylandrygholami,FTBcosmogholamilandry,scalarfieldTRW,ChaplyginpolyTRW}, because~using the same coframe/spin-connection pair and ansatz. However, the~current paper will allow us to verify, test, and determine the most suitable teleparallel $F(T)$ solutions on recent experimental data sets as performed in some studies~\cite{celiaBAO1,celiaBAO2,celiaBAO3}. We need, in~a near future and by using experimental data analysis approaches, to~compare the new solutions with BAO and redshift experimental data sets for determining the electromagnetic averaged contribution to universe models. This will allow us to select the most suitable classes of teleparallel $F(T)$ solutions useful for determining the averaged EMF contributions to the universe evolution models. {The $k=0$ case-based example used in Section~\ref{sect5} illustrates the feasibility of such a process for any teleparallel cosmological $F(T)$ solution with an electromagnetic source, including for a plasma-based universe model by the superposition principle. We can then repeat the same type of development for $k=\pm 1$ teleparallel cosmological $F(T)$ solutions.}

{There are some perspectives on future works with electromagnetic-based sources }for KS and static SS teleparallel spacetimes for new additional classes of solutions using the same process and coframe ansatz approaches as in refs.~\cite{staticscalarfieldSS,scalarfieldKS,nonvacSSpaper,nonvacKSpaper}. We also need to study in the near future the anti-deSitter (AdS) spacetimes in teleparallel gravity, including the electromagnetic influence on this spacetime structure. The~new electromagnetic teleparallel solutions found {in the paper will allow us to investigate some elaborated physical systems and processes in the teleparallel gravity context. These suggestions are really feasible in the near future and will contribute to the study of the charged AdS wormholes and BHs solutions in teleparallel gravity under a right teleparallel cosmological background.}


\vspace{6pt}

\section*{Abbreviations}
{
The following abbreviations are used in this manuscript:
\\

\noindent 
\begin{tabular}{l l}
AdS & Anti-deSitter\\
BH & Black Holes \\
CL & Conservation Law\\
DE & Dark Energy \\
{EM } & {Electromagnetic} \\
{EMF } & {Electromagnetic Field} \\
Eqn & Equation \\
FE & {Field Equation}\\
{GR} & {General Relativity }\\
KS & Kantowski--Sachs \\
{NGR} & {New General Relativity}\\
PF & Perfect Fluids\\
SF & Scalar Field\\
SS & Spherically Symmetric\\
TdS & Teleparallel deSitter \\
{TEGR} & {Teleparallel Equivalent of General Relativity} \\
TRW & Teleparallel Robertson--Walker\\
{$\Lambda$CDM } & {Lambda Cold Dark Matter}\\
\end{tabular}
}

\end{document}